\documentclass[aps,prl,twocolumn]{revtex4}
\usepackage{graphicx}
\usepackage[dvips,ps2pdf,colorlinks=true,urlcolor=blue]{hyperref}
\begin{document}
\title{Fermionic spin excitations in two and three-dimensional antiferromagnets}
\author{Zhihao Hao}
\author{Oleg Tchernyshyov}
\affiliation{Department of Physics and Astronomy, Johns Hopkins
University, Baltimore, MD, USA}

\begin{abstract}
Spin excitations in an ordered Heisenberg magnet are magnons---bosons with spin 1.  That may change when frustration and quantum fluctuations suppress the order and restore the spin-rotation symmetry.  We show that spin excitations in the $S=1/2$ Heisenberg antiferromagnet on kagome are spinons---fermions with spin $1/2$.  In the ground state the system can be described as a collection of small, heavy pairs of spinons with spin 0.  A magnetic excitation of lowest energy amounts to breaking up a pair into two spinons at a cost of $0.06 J$.
\end{abstract}

\maketitle

The Heisenberg antiferromagnet, described by the exchange
Hamiltonian
\begin{equation}
H = J\sum_{\langle ij \rangle} \mathbf S_i \cdot \mathbf S_j,
\label{eq:H-Heis}
\end{equation}
where $J>0$ is the exchange coupling and $\langle ij \rangle$
denotes a pair of neighboring sites, is a simple model of an
with realistic prototypes.  A bipartite antiferromagnet in three dimensions exhibits long-range spin order that breaks the global spin SU(2)
symmetry.  Low-energy excitations of such a magnet
are spin waves, whose quantization yields magnons---bosons
with spin 1 \cite{bec-review:2008}.  
In one dimension, spin order is disrupted by quantum fluctuations and the SU(2) symmetry is restored.  For spins of length
$S=1/2$, the excitations are spinons---quasiparticles with spin
1/2 \cite{faddeev:1981}.  A long-standing question 
is the existence of similarly unusual excitations in higher-dimensional
magnets where the symmetry of the ground state is restored by a combination of strong quantum fluctuations (small $S$) and geometrical
frustration (a non-bipartite lattice).  In 1973, Anderson
proposed a resonating valence-bond state for spins 1/2 on the triangular lattice
\cite{Anderson:1973}.  In this state, spins form singlet bonds, or quantum dimers \cite{PhysRevLett.61.2376}, with their neighbors.  Although the ground state on the triangular lattice turned out to be ordered, kagome and hyperkagome lattices have emerged as likely candidates for exotic quantum physics \cite{science.321.1306}.

\begin{figure}
\includegraphics[width=\columnwidth]{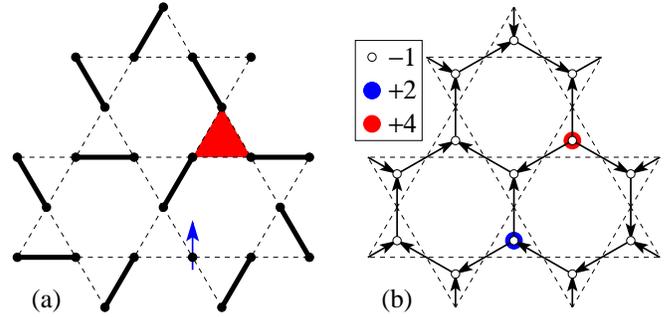}
\caption{(a) Kagome lattice and the major players: quantum dimers
(thick bonds), a defect triangle containing no dimer (red), and an
antikink spinon (blue arrow).  (b) The mapping to a compact U(1) gauge
theory on a honeycomb lattice.  Electric flux on links has strength
$\pm 1$ and is depicted as arrows in the manner of Elser and Zeng
\cite{PhysRevB.48.13647, PhysRevB.67.214413}.  The inset shows the
U(1) charges of the lattice sites and quasiparticles.}
\label{fig:kagome}
\end{figure}

The exchange energy (\ref{eq:H-Heis}) on kagome [Fig.~\ref{fig:kagome}(a)] is minimized when the total spin of every triangle attains its lowest allowed value \cite{PhysRevLett.62.2405}.  For $S=1/2$, that can be achieved by putting 
two spins of a triangle in a singlet state; the third spin is free to form a singlet on the adjacent triangle.  If a dimer could be placed on every triangle, we would construct a ground state with frozen singlets, a valence-bond solid.  That does not work: one in four triangles on kagome lack a dimer.  Quantum fluctuations induced by defect triangles make the ground state a nontrivial
superposition of valence-bond states.  Some recent theoretical works
\cite{PhysRevB.68.214415, singh:180407} lend support to a valence-bond crystal
proposed by Marston and Zeng \cite{marston:5962}, while others are consistent with a valence-bond liquid \cite{jiang:117203}.  Exact diagonalization studies \cite{EuroPhysJB.2.501} reveal a large number of low-lying singlet states, presumably associated with different dimer configurations.  Spin-1 excitations appear to have an energy gap of $0.05J$ to $0.10J$ \cite{EuroPhysJB.2.501, jiang:117203, singh:144415}.  

We propose a simple physical picture of the $S=1/2$
kagome antiferromagnet, in which the system is viewed as an ensemble
of spinons, fermions with spin 1/2.  
The antisymmetry of the many-body wavefunction can be traced to
quantum interference.  As two spinons are exchanged, they drag around
pairs of spins entangled in $S=0$ states.  In the end, 
the background singlet pairs return to the same positions.  However, an odd
number of them reverse their orientation, thus altering the sign of the many-body wavefunction.

The defect triangles turn out to be small, heavy pairs of spinons bound by exchange-mediated attraction.  The binding energy is $E_b = 0.06J$.  Spin-0
excitations are associated with the motion of pairs; their heavy
mass is reflected in the large density of singlet states at low
energies.  Spin-1 excitations correspond to breaking up a bound pair
into (nearly) free spinons with parallel spins.  Because spinon
pairs retain their individual character on kagome, the spin gap is
determined primarily by their binding energy, $\Delta \approx E_b =
0.06 J$.  This number is in line with the previous estimates of the
spin gap \cite{EuroPhysJB.2.501, singh:144415, jiang:117203}.

This picture is based on a study of two toy models on
lattices that share with kagome the triangular motif: the $\Delta$ chain \cite{PhysRevB.53.6393, PhysRevB.53.6401} and the
Husimi cactus, a tree-like variant of kagome
\cite{PhysRevB.48.13647, JPhysA.5.1541}.  Both systems have
dimerized ground states with exactly one singlet bond on every
triangle.  The $\Delta$ chain provides information about the
properties of individual spinons, while the Husimi cactus sheds
light on their quantum statistics and interactions.

\begin{figure}
\includegraphics[width=0.7\columnwidth]{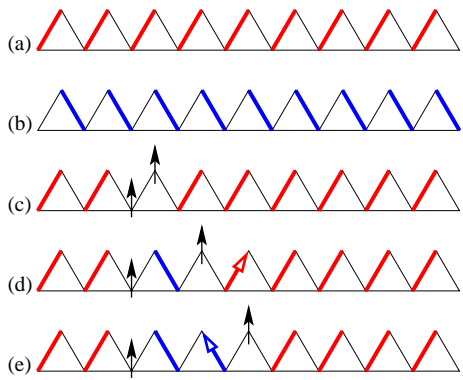}
\caption{Ground states $L$ (a) and $R$ (b) of the $\Delta$ chain.  A
local triplet excitation (c) decays into a localized kink and a
mobile antikink (d-e).  Arrows on the dimers illustrate the pivoting
rule (see text).} \label{fig:sawtooth}
\end{figure}

The $\Delta$ chain has two ground states, one with singlets on the
left bonds of the triangles ($L$), the other with singlets on the
right bonds ($R$), Fig.~\ref{fig:sawtooth}(a-b).  A local spin-1
excitation decays into two spinons---quasiparticles with spin 1/2,
Fig.~\ref{fig:sawtooth}(c-e).  They serve as domain walls and come
in two flavors \cite{PhysRevB.53.6393}: kinks interpolate between
state $L$ on the left and state $R$ on the right, antikinks do the
opposite.  A kink is fully localized and has \textit{zero} energy
cost: every triangle in its vicinity has a total spin 1/2 and is
thus in a state of lowest energy, which is of course a stationary
state.  One might think that zero excitation energy would lead to a 
proliferation of kinks, but that does not happen: kinks can only be 
created in pairs with antikinks, whose excitation energy is positive.  
That hints at the existence of a conserved charge, which will be defined 
later. 

Antikinks are mobile; their dynamics is described approximately as
hopping to the nearest triangle \cite{PhysRevB.53.6393}:
\begin{equation}
H |n \rangle
 = \frac{5J}{4}|n\rangle - \frac{J}{2} |n-1\rangle - \frac{J}{2} |n+1\rangle
 + \ldots
\label{eq:antikink-Delta}
\end{equation}
where $n$ is the triangle containing the antikink.  A Fourier
transform yields the quasiparticle dispersion $\epsilon(k) = 5J/4 -
J\cos{k} \approx J/4 + k^2/(2m)$ in the limit of low lattice
momenta; the antikink mass $m=1/J$.  A local spin-1 excitation
results in the creation of a kink and an antikink with a minimum
energy of $J/4$.  Terms omitted in Eq.~(\ref{eq:antikink-Delta})
create additional excitations by promoting the two singlets adjacent
to the antikink to triplets; alternatively, they can be viewed as
creation of an extra kink-antikink pair next to the antikink.  These
virtual excitations renormalize the antikink mass to $m \approx
1.16/J$ and lower the bottom of the antikink band to $\Delta =
0.219J$ \cite{supmat} in agreement with exact diagonalization \cite{PhysRevB.48.10552}.  Such
perturbations appear to be harmless and will be ignored.

We note that the sign of the hopping terms in
Eq.~(\ref{eq:antikink-Delta}) depends on the sign convention for 
quantum dimers \cite{raman:064413}: the $S=0$ state of two
spins is antisymmetric under exchange:
$|(i,j)\rangle \equiv (|\uparrow_i \downarrow_j\rangle -
|\downarrow_i \uparrow_j \rangle)/\sqrt{2} = -|(j,i)\rangle$.  When
necessary, we will depict a dimer state $|(i,j)\rangle$ as an arrow
pointing from $i$ to $j$.  The reversal of the arrow amounts to
multiplying the state wavefunction by $-1$.  A negative hopping amplitude is enforced by the \textit{pivoting rule:} when an antikink hops from one triangle to another, a quantum dimer moves in the opposite direction by pivoting
on the site shared by the triangles, Fig.~\ref{fig:sawtooth}(d-e).

\begin{figure}
\includegraphics[width=\columnwidth]{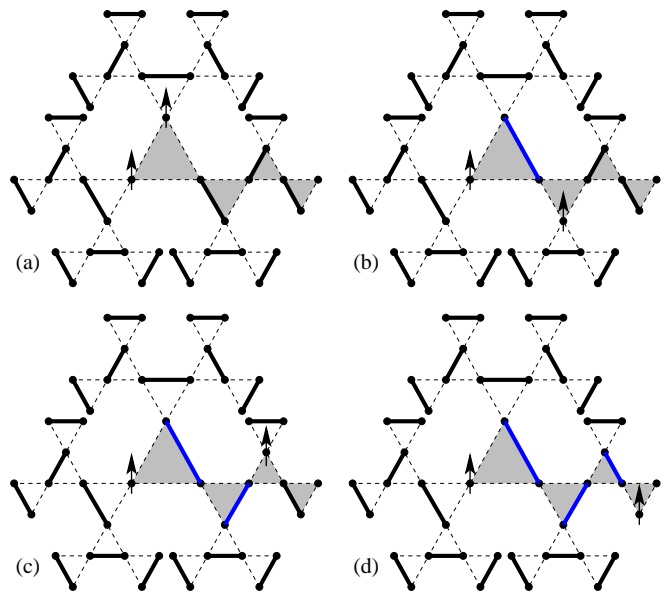}
\caption{A dimerized state on the Husimi cactus.  A local triplet
excitation (a) decays into a localized kink and a mobile antikink
(b-d).  Displaced valence bonds are shown in blue color.  Shaded
triangles in (d) mark the allowed path of the antikink.}
\label{fig:cactus-1}
\end{figure}

A ground state of the cactus has a
quantum dimer on every triangle \cite{JPhysA.5.1541}.  
Like on the $\Delta$ chain, a
local spin-1 excitation decays into a localized kink and a mobile
antikink propagating along a \textit{one-dimensional path}
(Fig.~\ref{fig:cactus-1}).  The kink has energy 0, while the
antikink propagates along the allowed path with a minimum energy of
$J/4$ and a mass $m=1/J$. (Again, we neglect virtual excitations 
in the vicinity of the antikink.)

\begin{figure}
\includegraphics[width=\columnwidth]{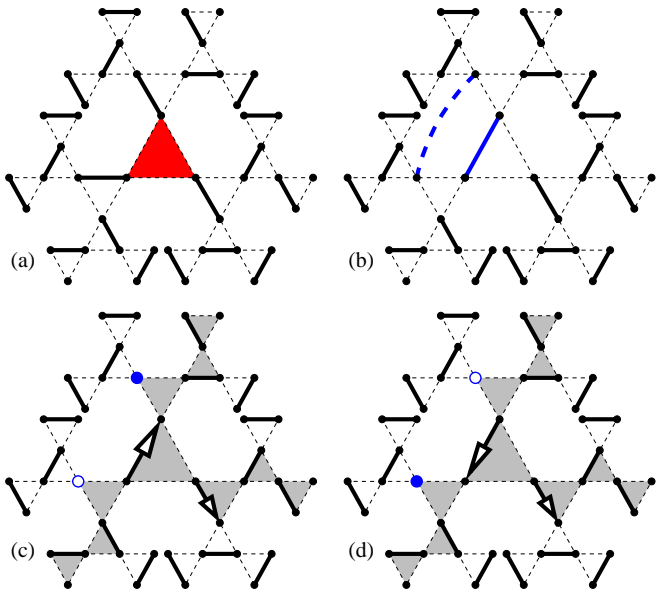}
\caption{A state with a single defect triangle (a) evolves into a
state with one longer-range bond (b), which can be viewed as two
antikinks, shown as open and filled blue dots (c), with a total spin
0.  State (d) is obtained from state (c) by an adiabatic exchange of
the antikinks.  Shaded triangles mark allowed paths of the
antikinks.  } \label{fig:cactus-0}
\end{figure}

Next we examine a single defect triangle on the 
cactus, Fig.~\ref{fig:cactus-0}(a).  The Hamiltonian connects this
state to a state with a singlet bond of a longer range \cite{PhysRevB.48.13647},
Fig.~\ref{fig:cactus-0}(b).  The longer-range bond can be viewed as
two antikinks with total spin 0.  By jumping to a
neighboring triangle, the antikinks are able to propagate along
\textit{three} branches of the cactus meeting at the original defect
triangle, Fig.~\ref{fig:cactus-0}(d).  Thus a defect
triangle is ``made" from two antikinks with total spin 0.
A state of two antikinks with a total spin 1 is also of interest to
us, so we will consider a generic case of two antikinks with some
specified spin projections, Fig.~\ref{fig:cactus-0}(c).

The Husimi cactus is a toy model in which
spinons have just enough freedom to exchange their locations by
hopping along the three branches.  That allows us to determine their
quantum statistics by examining the resulting Berry phase acquired by the
spinons.  The two-spinon wavefunction turns out to be 
antisymmetric under exchange, so antikinks are fermions.

We first sketch an informal argument along the lines of Arovas
\textit{et al.} \cite{PhysRevLett.53.722}.  Starting with the state
depicted in Fig.~\ref{fig:cactus-0}(c), we perform an adiabatic
exchange of the two antikinks using the pivoting rule.  We arrive at
the state shown in Fig.~\ref{fig:cactus-0}(d) with dimers in the
same positions but with one dimer reversed.  If we use the same
dimer basis to describe both states, the exchanged state acquires an
extra factor of $-1$.  The extra factor appears for any initial
positions of the antikinks.

The informal argument is supported by an explicit determination of
the ground state of two antikinks on the cactus \cite{supmat}.  We
find a nondegenerate, symmetric spatial wavefunction for total spin
$S=0$ and a doubly degenerate, antisymmetric spatial wavefunction
for $S=1$, as one expects for two fermions.  The ground state in the $S=1$ sector has energy $J/2$, twice the minimum energy of a free antikink. In contrast, two antikinks with $S=0$ form a bound state whose energy
lies $0.06 J$ below the bottom of the two-spinon continuum.
The bound state has a small size $\xi \approx 2.8$ lattice units of the $\Delta$ chain.

Curiously, antikinks in the $S=0$ ground state are bound not only to
each other, but also to the original defect triangle: a pair is
localized.  The localization is topological in origin: like on the
$\Delta$ chain, two antikinks cannot move through the same branch of
the cactus: an antikink may only be followed by a kink.

Let us discuss implications for the Heisenberg model on kagome.  The cactus can be viewed as a tree made of triangles. Extrapolating results from a tree to a periodic lattice (even with the same coordination number) is only warranted for those physical quantities that are not sensitive to the long-distance properties of the lattice.  For example, the band structure of a tree is quite different from that of a periodic lattice. Fortunately, the small size of a spinon pair means that its internal structure is determined by the local geometry of the lattice, so that the pair binding energy is expected to be roughly the same on the cactus, kagome, and hyperkagome.  

Both on kagome and hyperkagome one in four triangles carries a spinon pair.  We are now facing a many-body problem of fermionic spinons with a concentration of 1/3 per site.  The existence of spinons with Fermi-Dirac statistics in this and other frustrated two-dimensional magnets was conjectured previously \cite{marston:5962, PhysRevB.63.014413, hermele:224413} 
on the basis of a large-$N$ generalization of the Heisenberg model \cite{PhysRevB.37.3774}.  In these theories, spinons interact with an emergent lattice compact U(1) gauge field 
\cite{fradkin}.  We exploited an arrow representation of dimer coverings on kagome \cite{PhysRevB.48.13647, PhysRevB.67.214413} to define a fictitious U(1) gauge field living on links of a honeycomb lattice, Fig.~\ref{fig:kagome}(b).  The arrows depict a quantized electric field of unit strength.  Medial sites carry background charge $Q=-1$, which is neutralized on average by the charges of antikinks, $Q=+2$, and their pairs, $Q=+4$; kinks have charge $Q=-2$.  It is noteworthy that kinks and antikinks are indeed expected to carry a small real electric charge \cite{bulaevskii:024402}.

The gauge-theory connection sheds light on the origin of one-dimensional trajectories of spinons on kagome.  An object of charge $Q$ passing through a link increments the electric flux in it by $Q$, so antikinks can only move in the direction of electric flux, Fig.~\ref{fig:kagome}(b).  As a result, spinons are constrained to move along one-dimensional paths that may be infinite (as on the Husimi cactus) or may terminate in a loop.  
A finite length of an antikink trail $R$ would raise the triplet excitation energy of a pair by $E_c \sim \pi^2J/2R^2$ \cite{supmat}.  Since $R$ depends on the arrangement of dimers and may even vary from one spinon pair to the next, we only give an upper bound for the confinement energy.  On kagome, $R \geq 10$ so that $E_c \leq 0.05 J$.  Thus the spin gap $\Delta = E_b + E_c$ may vary from $0.06J$ to as much as $0.11J$, depending on the valence-bond pattern in the ground state.

Localization of antikink pairs can be traced to their interaction with the gauge field, whose electric flux is constrained to take on values $\pm 1$.  Two antikinks traveling along the same line would leave behind links with an unphysical electric flux $\pm 3$.  Therefore, in order to
move, the fermions making a pair must follow different paths.  The
tunneling amplitude is suppressed by a factor of order
$e^{-L/\xi}$, where $\xi \approx 2.8$ is the pair size and $L$ is
the length of the shortest loop on the medial lattice: $L = 6$ for
kagome, 10 for hyperkagome, and $\infty$ for the Husimi cactus.  
(The smallness of $e^{-L/\xi}$ also makes it possible to treat heavy spinon 
pairs as stationary objects on time scales relevant to the motion of light spinons.)  Singh and Huse \cite{singh:180407} found that the presence of other pairs increases the amplitude of pair tunneling.  Thus pairs form small localized clusters resonating around hexagons and preserving valence-bond order elsewhere.  

While our work agrees with the large-$N$ approaches \cite{marston:5962, PhysRevB.63.014413, hermele:224413, PhysRevB.37.3774} on the big picture of fermionic spinons interacting with a U(1) gauge field, there are important differences.  First, the spinons are not free: exchange-mediated attraction binds them into small bosonic pairs.  Second, the U(1) gauge field manifests itself as a quantized electric field, rather than a background magnetic flux, and frustrates the motion of spinon pairs.  The motion of spinons, which tends to scramble valence-bond order, is strongly suppressed as a result.  These observations lend support to the picture of a valence-bond crystal proposed by Marston and Zeng \cite{marston:5962, PhysRevB.68.214415, singh:180407}.  

Aside from inelastic neutron scattering, another way to break
spinon pairs is to apply a magnetic field strong enough to fully
polarize antikink spins, so that the magnetization reaches 1/3 of the
maximum value.  The liberated fermions will move at high speeds,
inducing strong fluctuations of the electric field and restoring
translational invariance.  A mean-field theory of spinons moving in
the background of the emergent magnetic flux may be a good starting
point for this partially polarized spin liquid.  That would be
interesting in light of recent numerical evidence for an
incompressible quantum spin liquid at 1/3 of full magnetization
\cite{JPSJ.70.3673, cabra:144420}.

We acknowledge helpful discussions with L. Balents, C. D. Batista, 
C. L. Henley, and V. Cvetkovic.  Research was supported in part by
the US DOE Grant No. DE-FG02-08ER46544.

\bibliography{kagome,quantum}

\end{document}